\documentclass[journal]{IEEEtran}

\usepackage{graphicx}
\usepackage{amsmath}
\usepackage{url}
\usepackage{hyperref}
\usepackage{multirow}

\title{ExDiff: A Framework for Simulating Diffusion Processes on Complex Networks with Explainable AI Integration}

\author{Annamaria Defilippo, Ugo Lomoio, Barbara Puccio, Pierangelo Veltri ~\IEEEmembership{Member,~IEEE}, and Pietro Hiram Guzzi ~\IEEEmembership{Senior Member,~IEEE}
\thanks{Annamaria Defilippo, Ugo Lomoio, Barbara Puccio, and Pietro Hiram Guzzi are with the Department of Surgical and Medical Sciences, Magna Graecua University of Catanzaro (e-mail: hguzzi.com)., Pierangelo Veltri is with DIMES, University of Calabria (pierangelo.veltri@unical.it)}}

\begin{document}

\maketitle

\begin{abstract}
Understanding and controlling diffusion processes in complex networks is critical across domains ranging from epidemiology to information science. Here, we present ExDiff, an interactive and modular computational framework that integrates network simulation, graph neural networks (GNNs), and explainable artificial intelligence (XAI) to model and interpret diffusion dynamics. ExDiff combines classical compartmental models with deep learning techniques to capture both the structural and temporal characteristics of diffusion across diverse network topologies. The framework features dedicated modules for network analysis, neural modeling, simulation, and interpretability, all accessible via an intuitive interface built on Google Colab. Through a case study of the Susceptible–Infectious–Recovered–Vaccinated–Dead (SIRVD) model, we demonstrate ExDiff’s capacity to simulate disease spread, evaluate intervention strategies, classify node states, and reveal the structural determinants of contagion through XAI techniques. By unifying simulation and interpretability, ExDiff provides a powerful, flexible, and accessible platform for studying diffusion phenomena in networked systems, enabling both methodological innovation and practical insight.
\end{abstract}


\IEEEpeerreviewmaketitle

\section{Introduction}

Understanding, predicting, and ultimately controlling the behaviour of complex systems remains a formidable challenge in scientific research and decision support. Computational models and infrastructures offer indispensable tools for bridging theory and application, enabling the simulation of dynamic phenomena across disciplines.

Diffusion processes on networks underpin numerous phenomena, ranging from information propagation in digital ecosystems to pathogen transmission in biological populations~\cite{hiram2022disease,guzzi2022pcn}. In technological domains, such analyses are critical for mapping innovation spread and detecting vulnerabilities within interconnected infrastructures~\cite{petrizzelli2022beyond}. Similarly, in ecological and environmental sciences, diffusion modelling elucidates species interactions and ecosystem robustness~\cite{de2016physics,abenavoli2014serum}. Simulations of these processes, grounded in a compact set of governing variables, allow researchers to both replicate emergent behaviours and predict the impact of perturbations, thereby providing a quantitative foundation for informed intervention~\cite{humphreys2002computational,milano2017extensive}.

Traditional approaches, such as compartmental models (e.g., SIR), offer tractable approximations for disease spread. However, they often neglect the underlying network structure of interactions. The COVID-19 pandemic highlighted the need for models that incorporate realistic contact topologies~\cite{petrizzelli2022beyond,guzzi2018survey}. This has motivated a shift towards contact-based models that couple compartmental dynamics with network theory, reflecting heterogeneous connectivity patterns observed in real-world systems~\cite{alguliyev2021graph,bryant2020modelling,karaivanov2020social,zaplotnik2020simulation,das2021analyzing,patil2021assessing,nassa2011comparative,cinaglia2}.

Despite advances in simulation methodologies, a significant gap persists in terms of accessible and extensible tools for non-expert users. Platforms such as NDLib~\cite{rossetti2018ndlib} provide web-based interfaces for diffusion studies but lack features critical for advanced network analysis, such as embedding-based inference and node-level classification.

To address these limitations, we introduce \textbf{ExDiff}~\footnote{Open-source software available at: \url{https://github.com/hguzzi/ExDiff}}, an explainable framework for simulating and analysing diffusion processes in complex networks. \textbf{ExDiff} (\textit{\textbf{Ex}plainable Graph Neural Network Framework for \textbf{Diff}usion Processes Modelling}) integrates classical compartmental models with Explainable Graph Neural Networks (GNNs)~\cite{gnn2009}, enabling interpretable predictions and scalable simulations across a wide array of network topologies.

The platform empowers users to simulate the spread of information, contagions, or behaviours across nodes in heterogeneous networks. This capacity is crucial for designing strategies to either amplify or suppress diffusion, tailored to specific application domains such as epidemiology, marketing, or cybersecurity. Compared to existing tools, \textbf{ExDiff} offers improved model interpretability, broader generalizability, and seamless integration with advanced deep learning modules.

Beyond simulation, \textbf{ExDiff} incorporates state-of-the-art deep learning capabilities~\cite{leskovec2010empirical,guzzi2022editorial,cho2013m}, supporting node embedding and classification tasks. This functionality enables the identification of influential or vulnerable nodes within diffusion pathways, thus informing targeted interventions. The fusion of explainable machine learning and network science positions \textbf{ExDiff} as a versatile platform for the rigorous study and control of complex diffusion dynamics.

The remainder of this paper is organized as follows. Section~\ref{sec:related} reviews related work. Section~\ref{sec:archi} presents the system architecture of \textbf{ExDiff}. Section~\ref{sec:simu} demonstrates its simulation capabilities, and Section~\ref{sec:casestudy} illustrates its application via a case study. 

\section{Related Work}
\label{sec:related}

\textbf{Graph Neural Networks.}
Graph Neural Networks (GNNs) have emerged as transformative tools for learning from graph-structured data, owing to their ability to iteratively propagate and aggregate information across node neighborhoods \cite{defilippo2024understanding}. Through message-passing mechanisms, GNNs capture both local connectivity patterns and broader network-level structures, enabling accurate performance in tasks such as node classification, link prediction, and community detection \cite{defilippo2024leveraging}.
Among GNN variants, Graph Convolutional Networks (GCNs) \cite{kipf2017semisupervised} represent a foundational architecture that extends convolutional operations to non-Euclidean domains. GCNs compute node embeddings via recursive feature aggregation from adjacent nodes, effectively encoding both intrinsic node attributes and structural context. Despite their success across a wide array of applications, the performance of GCNs on our diffusion modelling task proved limited. In particular, the phenomenon of over-smoothing—where node representations converge to indistinguishable vectors as layers increase—diminished their expressivity and utility in capturing critical heterogeneities.

The study of diffusion dynamics on graphs draws from a well-established tradition in network science and epidemiology, offering a unifying framework to model processes such as information spread, innovation adoption, and infectious disease transmission. The COVID-19 pandemic, with its unprecedented volume of empirical interaction data, has catalyzed a new wave of research into dynamic diffusion models that account for real-world complexity in contact structures.

\textbf{Compartmental models.}
Classical compartmental models such as Susceptible-Infectious-Recovered (SIR)~~\cite{kermack1927contribution} and its extension Susceptible-Infectious-Recovered-Vaccinated (SIRV) remain central to this endeavour. These models represent population dynamics through transitions between discrete health states and offer analytical tractability for evaluating intervention strategies. The inclusion of vaccination compartments in SIRV frameworks allows for a nuanced quantification of immunization effects on epidemic thresholds, transmission rates, and long-term containment \cite{plazas2021modeling}. Such extensions are crucial for assessing both spontaneous and policy-driven changes in population susceptibility.

\textbf{Graph-based modelling of diffusion processes.}
To capture the layered and evolving nature of human interactions, researchers have increasingly adopted multiplex and temporal network representations. Plazas et al. \cite{plazas2021modeling}, for instance, combined the SIR model with a multiplex network encoding household, workplace, and social contexts. This multilayer approach enabled a realistic simulation of disease progression and provided insights into the trade-offs between public health interventions and their economic consequences.

Temporal networks further enrich this modelling paradigm by incorporating the timing and order of interactions \cite{colizza2007reaction,milano2017extensive}. Humphries et al. \cite{humphries2021systematic} leveraged dynamic multiplex networks with time-varying edges to explore the temporal stability of epidemic control strategies. Their findings underscored the existence of critical parameter thresholds that delineate effective versus failed containment, emphasising the importance of temporal heterogeneity in determining diffusion outcomes.

Beyond epidemiological applications, diffusion mechanisms are integral to understanding how information, behaviours, and influence propagate in digital and social environments. A key focus in this area is the identification of super-spreaders—nodes with disproportionately high centrality or influence—that can accelerate or suppress the spread of content \cite{menczer2020first}. Targeting such nodes for intervention is a powerful strategy in domains ranging from public health messaging to the control of misinformation.

Taken together, the integration of advanced GNN architectures with epidemiological diffusion models, particularly within multiplex and temporal network frameworks, offers a powerful toolkit for analysing complex systems. These hybrid approaches enable researchers to move beyond static, homogeneous assumptions, illuminating how structural and temporal variability jointly shape diffusion trajectories. Such insights hold promise for guiding effective interventions across diverse domains, including healthcare, communication, and infrastructure resilience.

\section{System Architecture and Components}
\label{sec:archi}

\textbf{ExDiff} is a modular framework designed to simulate, analyze, and explain diffusion processes on complex networks by integrating graph theory, machine learning, and explainable AI components. Its architecture is composed of four core modules—\textit{Network Analysis}, \textit{Neural Networks (NN)}, \textit{Explainability}, and \textit{Simulation}—which are accessed and managed through an intuitive user interface.

\begin{figure}[h]
    \centering
    \includegraphics[width=0.55\linewidth]{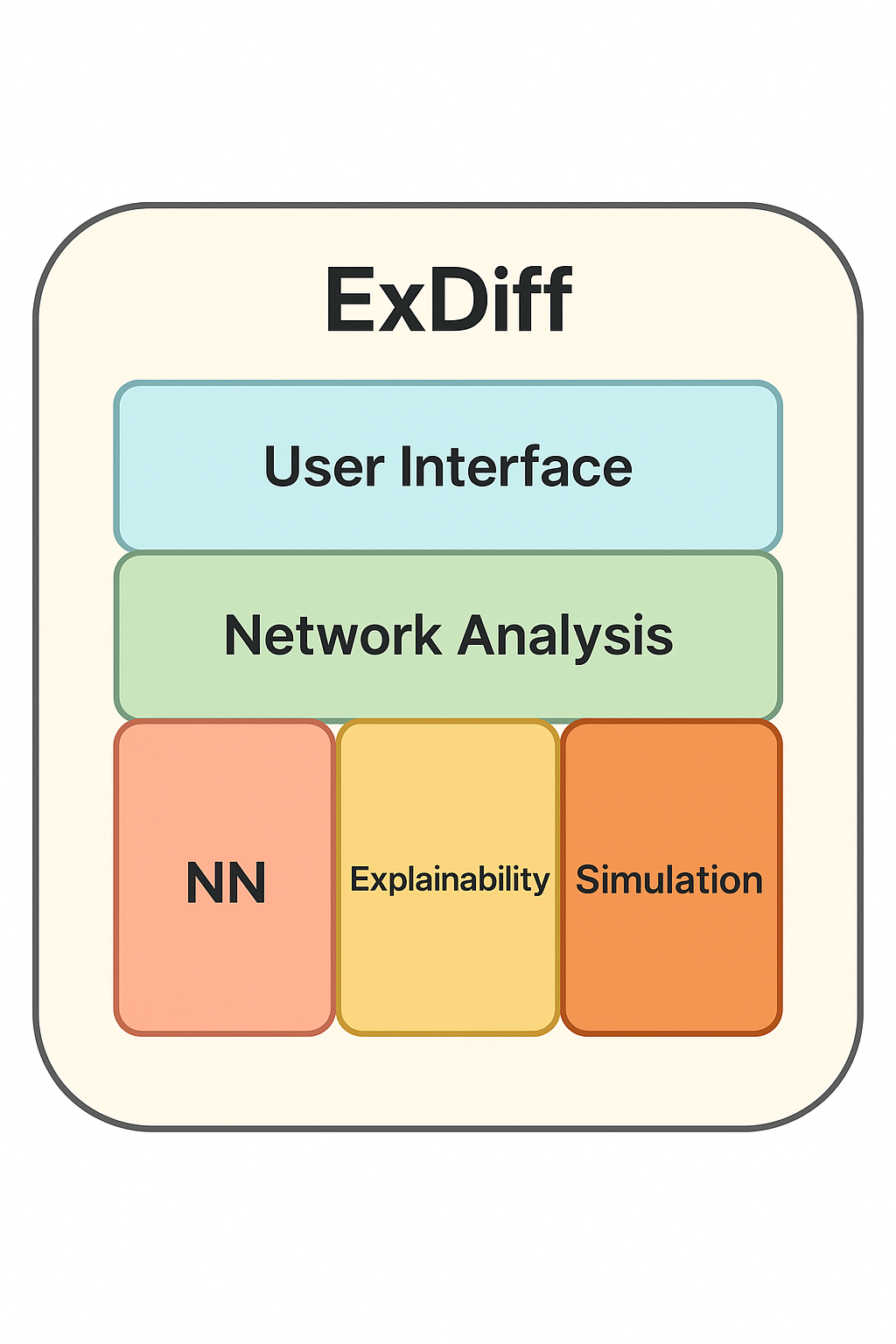}
    \caption{\textbf{Overview of the ExDiff architecture.} The system is organized in modular layers including a user interface, a central network analysis module, and three functional cores: neural network models (NN), explainability tools, and simulation engines. The visual representation is a placeholder and should be redesigned for publication.}
    \label{fig:architecture}
\end{figure}

\vspace{0.3cm}
\noindent \textbf{Network Analysis.} This module, built using the NetworkX library, underpins the structural representation and manipulation of graphs. It supports graph construction, computation of network metrics (e.g., betweenness centrality), and visualization of topologies. These analyses serve as preprocessing steps for downstream learning and simulation tasks.

\vspace{0.3cm}
\noindent \textbf{Neural Network (NN) Module.} Implemented in PyTorch and PyTorch Geometric, the NN module employs a $k$-GCN architecture to perform tasks such as node and edge classification, link prediction, and anomaly detection. It integrates GraphSAGE for inductive representation learning and PyGoD for detecting irregularities in graph structures. The module supports complete training and evaluation pipelines, allowing researchers to analyze dynamic graph-based phenomena.

\vspace{0.3cm}
\noindent \textbf{Explainability.} Transparency and interpretability are provided through Captum, a library for model interpretability in PyTorch. The module offers methods such as Integrated Gradients, Saliency Maps, and Layer-wise Relevance Propagation (LRP), enabling users to understand model decisions in terms of node features and connectivity. Both graphical and textual explanations are generated to support interpretation by domain experts.

\vspace{0.3cm}
\noindent \textbf{Simulation.} The simulation engine extends Menczer’s network dynamics class to model diffusion and behavioral evolution across time-varying topologies. It supports node- and edge-level temporal changes, scenario testing, and dynamic performance evaluation. A key feature is the ability to simulate vaccination strategies—including zero, random, and targeted immunization—by leveraging node rankings (e.g., betweenness centrality) to identify super-spreaders and evaluate intervention efficacy.

\vspace{0.3cm}
\noindent \textbf{User Interaction and Workflow.} Users interact with \textbf{ExDiff} via a streamlined Google Colab GUI. The interface allows for the selection and customization of network parameters, simulation settings, and visualization options. Once a graph is constructed, users can configure the SIRV model and initiate three parallel simulations corresponding to different vaccination strategies. The resulting diffusion curves are used to train the $k$-GCN model for predicting individual node states (Susceptible, Infected, Recovered, Vaccinated, Dead) at any given time point. Subsequently, the explainability module provides insights into the learned diffusion dynamics.

Overall, \textbf{ExDiff} offers a cohesive and extensible architecture for studying diffusion in complex systems. By integrating advanced graph neural networks with interpretable AI and network simulations, the framework supports a wide range of applications across epidemiology, information dynamics, and infrastructure resilience.



\section{Using ExDiff for Simulation }
\label{sec:simu}

The ExDiff user interface enable the user to perform a simulation on a network by setting many different parameters. 

First, Users can configure the desired network model from the supported models:

\begin{itemize}
    \item \textbf{Erdős–Rényi (ER)}: . \cite{menczer2020first}, introduced by Paul Erdős and Alfréd Rényi in 1959,  serves as a baseline for studying network structures. In this model, a graph is constructed by connecting each pair of $n$ nodes independently with a fixed probability $ p$. The ER model exhibits a binomial (or Poisson in the large  $n$ ) degree distribution and lacks the local clustering and community structure commonly found in real-world networks. Despite its simplicity, the ER model is powerful for analyzing phase transitions in connectivity, such as the emergence of a giant connected component as $p$ increases. It provides a mathematically tractable framework for understanding how random connections can give rise to complex global properties in networked systems.

    \item \textbf{Stochastic Block Model (SBM)}: The Stochastic Block Model (SBM) is a generative model for random graphs that captures community structure by dividing the set of $n$  nodes into $k$ distinct blocks or groups. The probability of an edge between any two nodes depends solely on the groups to which they belong, as specified by a $  k \times k$  block probability matrix. This allows SBM to model assortative (intra-group dense) or disassortative (inter-group dense) structures, making it highly suitable for studying networks with modular or clustered organization. Unlike the Erdős--Rényi model, which assumes uniform edge probability, SBM introduces heterogeneity and allows for the simulation of more realistic social, biological, and technological networks with latent group affiliations.

    \item \textbf{Random Geometric Graph (RGG):}The Random Geometric Graph (RGG) model generates networks by randomly placing $n$ nodes in a metric space—typically a unit square or cube—and connecting any pair of nodes whose Euclidean distance is less than a given threshold $r$. This spatial constraint results in graphs that exhibit high clustering and a strong locality of connections, properties often observed in wireless communication networks, sensor grids, and spatially embedded biological systems. The RGG model is particularly useful for analysing how geometry and physical proximity influence network topology and dynamics, as it naturally induces a spatial structure that differentiates it from non-embedded random graph models like Erdős--Rényi or SBM.

\end{itemize}

\subsection{Graph Visualization}
Network visualizations are created using the NetworkX and Matplotlib Python libraries, which offer advanced tools for rendering graph structures with precision and flexibility. The visualization process maps nodes and edges onto a two-dimensional space using layout algorithms that are tailored to the specific characteristics of the graph. For example, spring-force algorithms optimize the display of random networks, while geometric positioning is suited for spatially embedded models.

Visual encoding transforms abstract network properties into intuitive graphical elements. Node sizes, edge weights, and color schemes can represent various structural measures or dynamic states, such as centrality rankings or phases of disease propagation in epidemiological simulations. This visual mapping bridges the gap between numerical analysis and conceptual understanding, supporting both exploratory examination and detailed annotation of network behaviors.

Integrating real-time visualization within the simulation environment provides immediate visual feedback as parameters change, enabling researchers to observe how structural modifications affect network topology. The system also exports visualizations in publication-ready formats, including PNG and PDF files, ensuring consistency across research documentation and supporting reproducible workflows.

Beyond their analytical function, these visualizations serve as powerful communication tools that translate complex network phenomena into accessible visual narratives. An automated archival system creates a persistent visual record for each simulation run, establishing a comprehensive documentation system that facilitates cross-experiment comparisons and longitudinal analysis tracking.

\subsection{Export and Reusability}
The simulation platform enhances research continuity through Google Drive connectivity, offering persistent cloud storage for all experimental outputs, including network structures and visual representations. Each simulation generates a collection of artifacts, such as adjacency matrices and graphical outputs, along with records of temporal evolution when dynamics are involved.

Automated organizational systems arrange these outputs into structured directories, creating self-documenting records that maintain complete traceability and independence. This framework allows researchers to revisit and modify previous experiments without needing to reconstruct the entire pipeline.

The platform also emphasizes interoperability with standardized, human-readable output formats like JSON for parameters and CSV for graph data, enabling easy integration with external tools and collaborative environments.

\subsection{Explainable AI Integration}
ExDiff aims to incorporate XAI tools such as:
\begin{itemize}
    \item \textbf{SHAP values} to explain node feature contributions (e.g., centrality).
    \item \textbf{Counterfactual analysis} for structural interventions.
    \item \textbf{Causal inference} with Bayesian models.
\end{itemize}


\subsection{A Case Study}
\label{sec:casestudy}

To evaluate the practical utility and computational efficiency of the \textbf{ExDiff} framework, we conducted a case study using the Susceptible–Infectious–Recovered–Vaccinated–Dead (SIRVD) epidemiological model. The simulation was executed on the Google Colab platform, leveraging its accessibility for interactive experimentation and reproducibility. The primary goals of this case study were to assess the framework’s runtime performance, analyze node classification capabilities, and explore explainability features that illuminate the underlying diffusion dynamics.

The model parameters were set to reflect a realistic, though hypothetical, outbreak scenario. The initial population was divided among the five compartments: Susceptible (S), Infectious (I), Recovered (R), Vaccinated (V), and Dead (D). Key epidemiological parameters included the natural mortality rate of the human population, vaccination probability, exposure probabilities for vaccinated and unvaccinated individuals, infection and recovery probabilities, and pathogen-specific mortality rates. The contact network was constructed to emulate real-world interaction patterns, providing a rich substrate for diffusion processes.

Three vaccination strategies were simulated: (i) no vaccination, (ii) random vaccination, and (iii) targeted vaccination based on node centrality (e.g., betweenness centrality). These scenarios were designed to examine how different immunization strategies affect the dynamics of disease transmission and network resilience.

\begin{figure}[h]
\centering
\includegraphics[width=0.8\linewidth]{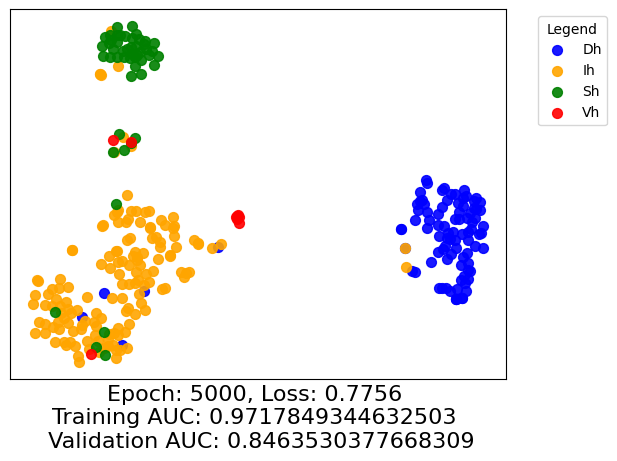}
\caption{\textbf{Latent space and node classification performance.} The $k$-GCN model effectively separates nodes into their epidemiological classes, demonstrating the network’s capacity to encode diffusion-relevant features.}
\label{fig:nodeclass}
\end{figure}

Node classification was performed using the $k$-GCN model integrated into \textbf{ExDiff}, trained to predict the compartmental state (S, I, R, V, or D) of each node at a given timestep. As shown in Figure~\ref{fig:nodeclass}, the model was capable of learning meaningful latent representations of the nodes, successfully separating them by class in the embedded latent space. These results indicate that the neural architecture effectively captures both structural and state-based features of the network.

\begin{figure}[h]
\centering
\includegraphics[width=0.8\linewidth]{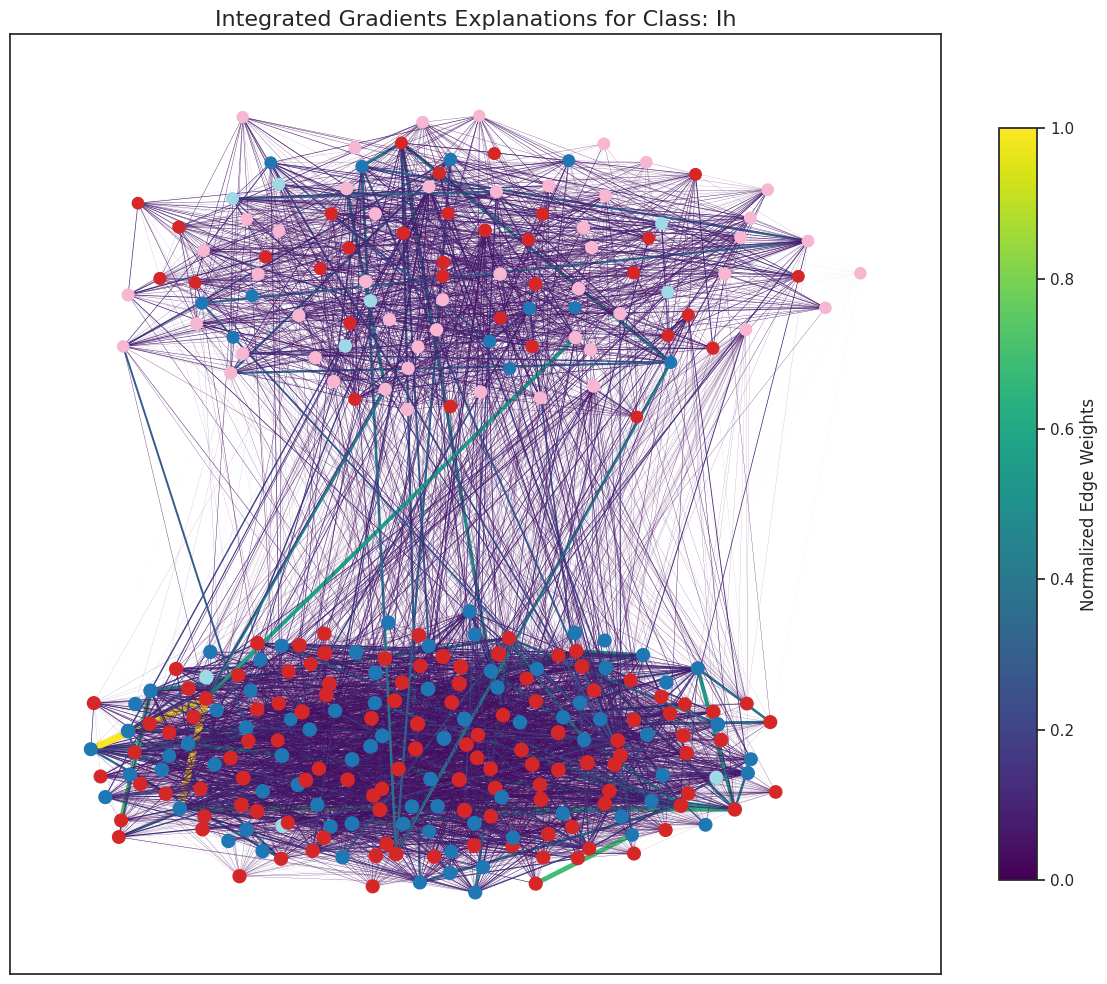}
\caption{\textbf{Explanation of diffusion dynamics.} Edges most relevant to disease transmission are highlighted. Node colors represent their compartmental status: blue (Dead, D\textsubscript{h}), pink (Susceptible, S\textsubscript{h}), cyan (Vaccinated, V\textsubscript{h}), and red (Infectious, I\textsubscript{h}).}
\label{fig:exp}
\end{figure}

To complement the classification analysis, the explainability module was employed to interpret the model's predictions. Using methods such as Integrated Gradients and Layer-wise Relevance Propagation, the framework identified critical nodes and edges contributing to the spread of infection. Figure~\ref{fig:exp} illustrates the most influential connections for disease transmission, with node coloring reflecting their epidemiological status: blue (Dead, D\textsubscript{h}), pink (Susceptible, S\textsubscript{h}), cyan (Vaccinated, V\textsubscript{h}), and red (Infectious, I\textsubscript{h}). This visual representation provides insight into how infection propagates through the network and which individuals or connections are most responsible for transmission.

\section{Conclusion}
\label{sec:conclusion}
We have presented \textit{ExDiff}, a Python-based simulation framework integrating graph-theoretic modeling with explainability tools. ExDiff enables accessible, reproducible experimentation with potential applications in public health, social networks, and computational science.


\bibliographystyle{unsrt}
\bibliography{biblio}
\end{document}